\documentclass[aps,pra,showpacs,twocolumn,superscriptaddress]{revtex4}

\usepackage{graphicx}
\usepackage{dcolumn}
\usepackage{bm}
\usepackage{amsfonts}
\usepackage{amsmath}
\usepackage{color}

\newcommand{\beq}{\begin{equation}}
\newcommand{\eeq}{\end{equation}}
\newcommand{\bea}[1]{\begin{equation}\begin{array}{#1}}
\newcommand{\eea}{\end{array}\end{equation}}
\newcommand{\beqn}{\begin{eqnarray}}
\newcommand{\eeqn}{\end{eqnarray}}

\begin{document}

\title{Environment-induced entanglement with a single photon}

\author{M. Hor-Meyll}
\email{malena@if.ufrj.br}
\affiliation{Instituto de F\'{\i}sica, Universidade Federal do Rio
de Janeiro, Caixa Postal 68528, Rio de Janeiro, RJ 21941-972,
Brazil}

\author{A. Auyuanet}
\affiliation{Instituto de F\'{\i}sica, Universidade Federal do Rio de
Janeiro, Caixa Postal 68528, Rio de Janeiro, RJ 21941-972, Brazil}

\author{C. V. S. Borges}
\affiliation{Instituto de F\'{\i}sica, Universidade Federal Fluminense, Niter\'oi, RJ 24210-340,
Brazil}

\author{A. Arag\~ao}
\affiliation{Instituto de F\'{\i}sica, Universidade Federal do Rio de
Janeiro, Caixa Postal 68528, Rio de Janeiro, RJ 21941-972, Brazil}

\author{J. A. O.  Huguenin}
\affiliation{Departamento de Ci\^encias Exatas, PUVR-UFF, Volta Redonda, RJ  27255-125, Brazil}

\author{A. Z. Khoury}
\affiliation{Instituto de F\'{\i}sica, Universidade Federal Fluminense, Niter\'oi, RJ 24210-340,
Brazil}

\author{L. Davidovich}
\affiliation{Instituto de F\'{\i}sica, Universidade Federal do Rio de
Janeiro, Caixa Postal 68528, Rio de Janeiro, RJ 21941-972, Brazil}

\begin{abstract}
We propose an  all-optical setup, which couples different degrees of freedom of a single photon, to investigate entanglement generation by a common environment. The two qubits are represented by the photon polarization and Hermite-Gauss transverse modes, while the environment corresponds to the photon path. For an initially two-qubit separable state, the increase of entanglement is analyzed, as the probability of an environment-induced transition ranges from zero to one. An entanglement witness that is invariant throughout the evolution of the system yields a direct measurement of the concurrence of the two-qubit state.   \end{abstract}


\maketitle

\section{INTRODUCTION}
\label{sec:introduction}

The investigation of the impact of the environment on multi-partite entangled systems has brought out some subtle aspects of entanglement. Its time-dependent behavior can be much different from that of the decay of the populations of the individual parts, or of the coherence between them. In fact, it has been demonstrated theoretically \cite{ karol, simon0, diosi, dodd, dur, yu1, carvalho:230501,mintert, hein, fine:153105, santos:040305, yu:140403,  carvalho-2007,aolita} and experimentally \cite{almeida07,kimble07,alejo} that entanglement may vanish much before coherence disappears. 

Linear optical devices have been shown to be a powerful tool for the investigation of this phenomenon \cite{almeida07,aiello07,alejo}. By associating the polarization degrees of freedom with qubits, and the momentum degrees of freedom with the environment, one has been able to realize experiments that probe the dynamics of entanglement of qubit systems interacting independently with individual environments, which, as an added bonus, can be tailored at will and continuously monitored.  In particular,  ``entanglement sudden death" \cite{yu:140403} was demonstrated \cite{almeida07}, and the evolution of a qubit under continuous monitoring of the environment was investigated \cite{alejo}. 

The assumption of independent environments requires that the individual systems be sufficiently separated; for atoms interacting with the reservoir of electromagnetic field modes, this would mean that they should be much farther apart than the wavelength of the resonant radiation. If this condition is not met,  the interaction of the parts with a common environment may give rise to collective effects.  In this case, the direct interaction between the individual systems may substantially affect the collective behavior of the system, as has been extensively discussed within the realm of superradiance \cite{dicke, haroche}: thus, for instance, when atoms get closer together, dipole interactions should be taken into account. 

In some situations, however, the direct interaction between the individual components of the system does not mask the effects of the interaction with a common environment. This is the case, for instance, in ion trap experiments, where neighboring ions may be subject to a common dephasing process. Under these conditions, it is possible to protect the system against decoherence by defining logical qubits in decoherence-free subspaces \cite{palma,barenco,duan,zanardi,lidar1,guo,lidar,bacon,aolita1}, which have been experimentally demonstrated with twin-photon beams \cite{kwiatDFS}, nuclear magnetic resonance \cite{viola,fortunato}, and trapped ions \cite{kielpinski,roos,langer,haffner}. 

A common environment may entangle initially separable states \cite{braun,kimheat,jakobczyk,schneider,luodecoprod,davies,derkacz,jakojamspont,benatti,ficekspontaneous2,ficekdark}, even when it is in a thermal state \cite{braun,kimheat}. Two problems arise however in possible experimental demonstrations of this property: this effect can be masked by the direct interaction between the individual parts of the system and it may be difficult to prepare the qubits in arbitrary individual states due to their proximity. 

Here we show that an all-optical setup that combines different degrees of freedom of a single photon can be used to investigate the role of a common environment in the generation of entanglement between two non-interacting qubits, which can be prepared in independent individual states at will. Polarization and transverse mode degrees of freedom of the photon stand for the two qubits, and the photon path is the environment. Coupling between the qubits and the environment is achieved with linear optics devices. 

The use of multiple degrees of freedom of photons, while not leading to scalable quantum computation \cite{blume-kohout02}, has allowed the study of  basic quantum algorithms \cite{cerf98,oliveira05,walborn05c}, quantum teleportation \cite{boschi98}, entanglement purification \cite{pan03}, improved Bell-state analysis \cite{kwiat98a,walborn03b,walborn03c,kwiat07}, creation of high-dimensional entanglement \cite{barreiro05}, demonstration of direct methods for measuring entanglement~\cite{walborn06b,walborn07a}, engineering of mixed states through decoherence \cite{peters04,aiello07}, quantum key distribution \cite{aolita:100501,souza:032345}, measurement of a topological phase \cite{souza:160401} and the investigation of the environment-induced dynamics of entangled systems \cite{almeida07,alejo}. 

Our proposed setup is divided into two parts. The first one describes the action of a common environment acting on the two qubits; the second analyses the generated state through the measurement of an entanglement witness, which has a peculiar and very useful property: it is time-independent, so the same setup can be used, as the environment-induced transition probability ranges from zero (corresponding to the initial time) to one. 

This paper is organized as follows. In Section \ref{sec:theoretical model} we present the master equation that describes the interaction of an ensemble of qubits with a common environment. The corresponding unitary map, essential for building the experiment, is introduced in Section \ref{sec:unitary map}. In Section \ref{sec:entanglement measurement} we present the single observable that allows for a direct measurement of entanglement, thus avoiding the cumbersome process of tomography \cite{kwiat-tomo}. The proposed experiment, using a linear optical setup, which leads to entanglement creation, plus the measurement circuit to quantify it, are  presented in Section \ref{sec:experiment}. In Section \ref{sec:generalization of the model} we show that the same optical circuit can be used to probe the coherent interaction between two two-level systems and a single-mode environment. The conclusions are presented in Section \ref{sec:conclusion}. In the Appendix we derive the Kraus operators from the master equation describing the interaction of two qubits with a common environment.

\section{THEORETICAL MODEL}
\label{sec:theoretical model}
We consider a system of $N$ identical qubits interacting with a common zero-temperature environment. The state of each qubit is represented in the basis $\{|e\rangle,|g\rangle\}$, where we take $|e\rangle$ to be the excited state and $|g\rangle$ the ground state. The state $|e\rangle$ decays into the state $|g\rangle$, producing one excitation in the environment. We may associate to qubit $i$  the Pauli operator $S^{z}_{i}=|e_{i}\rangle \langle e_{i}|-|g_{i}\rangle \langle g_{i}|$ and the ladder operators $S^{+}_{i}=|e_{i}\rangle \langle g_{i}|$ and  $S^{-}_{i}=|g_{i}\rangle \langle e_{i}|$. The collective interaction between the qubits and the infinite modes of a bosonic reservoir is described in the rotating-wave approximation by the Hamiltonian
\begin{eqnarray}
H= \sum_{i=1}^N \hbar \omega S_{i}^{z}  +\sum_{\ell}
\hbar \omega_{\ell}\left(a^{\dag}_{\ell}a_{\ell}+1/2\right)\nonumber \\
 - i\hbar
\sum_{\ell} \sum_{i=1}^{N}
\left[{g}_{\ell}\left(S_{i}^{+}a_{\ell}-S_i^{-}a_{\ell}^{\dagger}\right)\right],
\label{eq1}
\end{eqnarray}
where  ${g}_{\ell}$ is the coupling constant (taken to be real), $\omega$ is the transition frequency between the two qubit levels, $a_{\ell}$ and $a^{\dag}_{\ell}$ are the annihilation and creation operators corresponding to mode $\ell$ with frequency $\omega_{\ell}$. Here we have neglected  the spatial dependence of the coupling. 

The interaction of the qubits with the common environment is mediated by the collective operators $\displaystyle S^{\pm}= \sum_{i=1}^N S_{i}^{\pm}$.  The dynamical evolution of the qubit system can be described by a master equation, which in the interaction picture, under the usual Born-Markov approximation, and neglecting the direct interaction between the qubits, is given by \cite{haroche,ficek}:
\begin{eqnarray}
\frac{\partial {\rho}_{\mathcal{S}}(t)}{\partial t}=
\frac{\Gamma}{2}[2S^{-}\rho_{\mathcal{S}}(t)S^{+}
-\rho_{\mathcal{S}}(t)S^{+}
S^{-}-
S^{+}S^{-}\rho_{\mathcal{S}}(t)],\nonumber\\
\label{lindblad}
\end{eqnarray}
where $\rho_{\mathcal{S}}(t)$ is the density matrix of the system and the spontaneous decay rate $\Gamma$ is proportional to the square of the coupling constant evaluated at the transition frequency $\omega$. Thus, for instance, for a two-level atom interacting with the electromagnetic field, one may take $\ell\equiv\{\vec k,s\}$, where $\vec k$ is the wave vector corresponding to a plane-wave mode and $s$ is the corresponding polarization index. Then, $g_{\ell}=\left( \omega_{k}/2 \epsilon_{0}\hbar V\right)^{\frac{1}{2}}\vec{e}_{\vec{k}{s}}\cdot\vec{\mu}$, where $V$ is the quantization volume, $\vec{e}_{\vec{k}{s}}$ is a polarization vector, and $\vec\mu$ is the transition dipole moment. Then $\Gamma=
\omega^{3}\mu^{2}/3\pi\epsilon_{0}\hbar c^{3}$. In the following, we specialize our discussion to two-qubit systems. 

\section{UNITARY MAP AND KRAUS OPERATORS}
\label{sec:unitary map}

The interaction of a system ${\cal S}$ with an environment ${\cal E}$ can be described in terms of a unitary evolution encompassing the system and the environment together:
\begin{eqnarray}
|\phi^{i}\rangle_{\mathcal{S}}|0\rangle_{\mathcal{E}}&\rightarrow&\sum_{\mu}M_{\mu}|\phi ^{i}\rangle_{\mathcal{S}}|\mu\rangle_{\mathcal{E}},
\label{unitarymap}
\end{eqnarray}
where $|\phi^{i}\rangle_{\mathcal{S}}$ $ (i=1...4)$ are orthogonal states of the
system, $|\mu\rangle_{\mathcal{E}}$ are orthogonal states of the
environment, and $M_{\mu}$ are the so-called Kraus operators \cite{kraus}, acting only on the states of the system ${\cal S}$, and satisfying
\begin{eqnarray}
\sum_{\mu}M^{\dag}_{\mu}M_{\mu}=1. 
\label{krausnorm}
\end{eqnarray}

As stressed in \cite{alejo}, the map (\ref{unitarymap}) is actually more general than the master equation (\ref{lindblad}), which is obtained from it, when the environment has many degrees of freedom, under Markovian and differentiability assumptions \cite{lindblad,preskill}. 

The Kraus operators corresponding to Eq.~(\ref{lindblad}) are obtained in the Appendix, using  the Choi matrix formalism \cite{choi,havel}.
Substituting the resulting Kraus operators (\ref{kraus0}-\ref{kraus2}) in Eq.~(\ref{unitarymap}),  we get, in the collective basis $\{|0,0\rangle\equiv\frac{1}{\sqrt{2}}(|eg\rangle-|ge\rangle), |1,1\rangle\equiv|ee\rangle, |1,0\rangle\equiv \frac{1}{\sqrt{2}}(|eg\rangle+|ge\rangle), |1,-1\rangle\equiv|gg\rangle\}$:
\begin{eqnarray}
|1,1\rangle_{\mathcal{S}}|0\rangle_{\mathcal{E}}&\rightarrow& A|1,1\rangle_{\mathcal{S}}|0\rangle_{\mathcal{E}}+B|1,0\rangle_{\mathcal{S}}|1_{A}\rangle_{\mathcal{E}}+\nonumber\\
&&D|1,0\rangle_{\mathcal{S}}|1_{B}\rangle_{\mathcal{E}}+F |1,-1\rangle_{\mathcal{S}}|2\rangle_{\mathcal{E}}\,,\nonumber\\
|1,0\rangle_{\mathcal{S}}|0\rangle_{\mathcal{E}}&\rightarrow& A
|1,0\rangle_{\mathcal{S}}|0\rangle_{\mathcal{E}}+C |1,-1\rangle_{\mathcal{S}}|1_{A}\rangle_{\mathcal{E}} +\nonumber\\
&& E|1,-1\rangle_{\mathcal{S}}|1_{B}\rangle_{\mathcal{E}} \,, \nonumber \\
|1,-1\rangle_{\mathcal{S}}|0\rangle_{\mathcal{E}}&\rightarrow&
|1,-1\rangle_{\mathcal{S}}|0\rangle_{\mathcal{E}} \,, \nonumber \\
|0,0\rangle_{\mathcal{S}}|0\rangle_{\mathcal{E}}&\rightarrow&
|0,0\rangle_{\mathcal{S}}|0\rangle_{\mathcal{E}}\,, \label{mapa0}
\end{eqnarray}
where $A, B, \cdots F$ are time-dependent coefficients explicitly calculated in the Appendix and the states of the environment are labelled so as to highlight their physical meaning in terms of the number of excitations, i.e., $|0\rangle_{\mathcal{E}}$ corresponds to the vacuum state, $|1_{A}\rangle_{\mathcal{E}}$ and  $|1_{B}\rangle_{\mathcal{E}}$ to states with just a single excitation and $|2\rangle_{\mathcal{E}}$ to a state with two excitations in the environment. This set of equations has an immediate physical interpretation: each coefficient represents the probability amplitude of transition between the qubit states as a function of time, either the qubits emitting excitations into the environment, which is initially in the vacuum state, or exchanging excitations between them, or yet not emitting any excitation at all, when both system and environment remain in the same state. For instance, in the first line,
where both qubits are initially in the excited state $|1,1\rangle_{\mathcal{S}}$, we have a probability amplitude A that both qubits remain in the excited state, with no excitation being emitted into the environment,
which remains in the vacuum state $|0\rangle_{\mathcal{E}}$. On the other hand, there is a probability
amplitude B(D) that a single qubit decays, the state of the system going to
a superposition of $|eg\rangle_{\mathcal{S}}$ and $|ge\rangle_{\mathcal{S}}$ with just one excitation being
transferred to the environment, which goes to state $|1_{A}\rangle_{\mathcal{E}}$ ($|1_{B}\rangle_{\mathcal{E}}$). Finally there is a probability amplitude F that both qubits decay to $|1,-1\rangle_{\mathcal{S}}$ emitting two photons into the environment, whose state becomes
$|2\rangle_{\mathcal{E}}$. Similar reasonings can be applied to the other lines of the map. 

We note, in Eq.~(\ref{mapa0}), the appearance of two orthogonal environment states with just a single excitation, $|1_{A}\rangle_{\mathcal{E}}$ and $|1_{B}\rangle_{\mathcal{E}}$. This can be understood physically by rewriting Eq.~(\ref{mapa0}) as:
\begin{eqnarray}
|1,1\rangle_{\mathcal{S}}|0\rangle_{\mathcal{E}}&\rightarrow& A|1,1\rangle_{\mathcal{S}}|0\rangle_{\mathcal{E}}+G|1,0\rangle_{\mathcal{S}}|1_{1,1}\rangle_{\mathcal{E}}+\nonumber\\
&& F |1,-1\rangle_{\mathcal{S}}|2\rangle_{\mathcal{E}}\,, \nonumber\\
|1,0\rangle_{\mathcal{S}}|0\rangle_{\mathcal{E}}&\rightarrow& A
|1,0\rangle_{\mathcal{S}}|0\rangle_{\mathcal{E}}+H|1,-1\rangle_{\mathcal{S}}|1_{1,0}\rangle_{\mathcal{E}} \,,\nonumber \\
|1,-1\rangle_{\mathcal{S}}|0\rangle_{\mathcal{E}}&\rightarrow&
|1,-1\rangle_{\mathcal{S}}|0\rangle_{\mathcal{E}} \,,\nonumber \\
|0,0\rangle_{\mathcal{S}}|0\rangle_{\mathcal{E}}&\rightarrow&
|0,0\rangle_{\mathcal{S}}|0\rangle_{\mathcal{E}}, \label{mapa1}
\end{eqnarray}
where $|1_{1,1}\rangle_{\mathcal{E}}$ and $|1_{1,0}\rangle_{\mathcal{E}}$ are non-orthogonal states of the environment with a single excitation given by:
\begin{eqnarray*}
|1_{1,1}\rangle_{\mathcal{E}}&\equiv&\left(\frac{B}{G}|1_{A}\rangle_{\mathcal{E}}+\frac{D}{G}|1_{B}\rangle_{\mathcal{E}}\right)\,, \\
G&\equiv&\sqrt{B^{2}+D^{2}}\,,\\
|1_{1,0}\rangle_{\mathcal{E}}&\equiv& \left(\frac{C}{H}|1_{A}\rangle_{\mathcal{E}}+\frac{E}{H}|1_{B}\rangle_{\mathcal{E}}\right)\,, \\
H&\equiv&\sqrt{C^{2}+E^{2}}.
\label{grouping}
\end{eqnarray*}
The explanation for the presence of two different states with a single excitation in the first and second lines of Eq.~(\ref{mapa1}) stems from the consideration of the photon emission process for the initial states $|1,1\rangle$ and $|1,0\rangle$, respectively. The rate of photon emission for the two qubits is given by \cite{haroche}:
\begin{eqnarray}
W_{2}=\Gamma\langle S^{+}S^{-}\rangle.
\label{photonemission}
\end{eqnarray}

Calculating this rate for the initial states $|1,1\rangle$ and $|1,0\rangle$ we get:
\begin{eqnarray}
W^{1,1}_{2}=\Gamma\langle 1,1| S^{+}S^{-} |1,1\rangle=2\Gamma,\nonumber\\
W^{1,0}_{2}=\Gamma\langle 1,0| S^{+}S^{-} |1,0\rangle=2\Gamma.
\label{staterates}
\end{eqnarray}

The rate corresponding to $|1,1\rangle$ is exactly the one we would expect from two independent qubits. This implies that in this case the first photon emitted would have a linewitdh of $\Gamma$. The decay rate of the single excited qubit in the state $|1,0\rangle$ is twice that of each qubit in $|1,1\rangle$, so this state is superradiant and the linewidth of the emitted photon is expected to be $2\Gamma$. As we have two different linewidths for the photons emitted from the initial states $|1,1\rangle$ and $|1,0\rangle$,  this explains why we have two different environment states corresponding to a single excitation, $|1_{1,1}\rangle_{\mathcal{E}}$ and $|1_{1,0}\rangle_{\mathcal{E}}$ in the first and the second line of Eq.~(\ref{mapa1}), respectively.

In terms of the computational basis $\{|ee\rangle, |eg\rangle, |ge\rangle, |gg\rangle\}$, Eq.~(\ref{mapa0}) can be rewritten as
\begin{eqnarray}
|ee\rangle_{\mathcal{S}}|0\rangle_{\mathcal{E}}&\rightarrow&
M|ee\rangle_{\mathcal{S}}|0\rangle_{\mathcal{E}}+P (|eg\rangle+ |ge\rangle)_{\mathcal{S}}|1_{ee}\rangle_{\mathcal{E}}+\nonumber \\
&&N |gg\rangle_{\mathcal{S}}|2\rangle_{\mathcal{E}}\,,\nonumber \\
|eg\rangle_{\mathcal{S}}|0\rangle_{\mathcal{E}}&\rightarrow& Q
|eg\rangle_{\mathcal{S}}|0\rangle_{\mathcal{E}}+ R |ge\rangle_{\mathcal{S}}|0\rangle_{\mathcal{E}} + S|gg\rangle_{\mathcal{S}}|1_{eg}\rangle_{\mathcal{E}}\,,\nonumber\\
|ge\rangle_{\mathcal{S}}|0\rangle_{\mathcal{E}}&\rightarrow&
Q |ge\rangle_{\mathcal{S}}|0\rangle_{\mathcal{E}}+ R |eg\rangle_{\mathcal{S}}|0\rangle_{\mathcal{E}} + S|gg\rangle_{\mathcal{S}}|1_{eg}\rangle_{\mathcal{E}}\,,
\nonumber\\
|gg\rangle_{\mathcal{S}}|0\rangle_{\mathcal{E}}&\rightarrow&
|gg\rangle_{\mathcal{S}}|0\rangle_{\mathcal{E}}, \label{mapa2}
\end{eqnarray} \\
where $M= e^{-\Gamma t}$, $P= \sqrt{\Gamma t e^{-2\Gamma t}}$, $N=\sqrt{1-e^{-2\Gamma t}-2\Gamma t  e^{-2\Gamma t}}$, $Q= \frac{e^{-\Gamma t}+1}{2}$, $R= \frac{e^{-\Gamma t}-1}{2}$, $S= \sqrt{\frac{1-e^{-2\Gamma t}}{2}}$. The new single-excitation states $|1_{ee}\rangle_{\mathcal{E}}
$ and $|1_{eg}\rangle_{\mathcal{E}}
$ are related to orthogonal states of the environment $|1_{A}\rangle_{\mathcal{E}}
$ and $|1_{B}\rangle_{\mathcal{E}}
$ by:
\begin{eqnarray}
|1_{ee}\rangle_{\mathcal{E}}&\equiv&\left(\frac{X}{P}|1_{A}\rangle_{\mathcal{E}}+\frac{Y}{P}|1_{B}\rangle_{\mathcal{E}}\right)\,, \nonumber\\
|1_{eg}\rangle_{\mathcal{E}}&\equiv& \left(\frac{Z}{S}|1_{A}\rangle_{\mathcal{E}}+\frac{W}{S}|1_{B}\rangle_{\mathcal{E}}\right), 
\label{eq16}
\end{eqnarray}
where $X$, $Y$, $Z$ and $W$ are time-dependent coefficients defined in the Appendix.

As shown in Section \ref{sec:experiment}, Eq.~(\ref{mapa2}) leads to a simple linear optics demonstration of environment-induced entanglement.

\section{ENTANGLEMENT GENERATION AND MEASUREMENT}
\label{sec:entanglement measurement}

A detailed analysis of the generation of entanglement between two atoms via spontaneous emission, with dipole-dipole interaction, was presented by Ficek and Tana\'s~\cite{ficekspontaneous2,ficekdark}.  Here we consider the special case in which the separation between the atoms is much smaller than the typical wavelength of the emitted radiation, and there is no dipole-dipole interaction. Our expressions coincide with those in Refs.~\cite{ficekspontaneous2,ficekdark} in the proper limit.

For the quantification of entanglement we use the  \textit{concurrence} \cite{wootters}, defined as:
 \begin{equation}
{\cal C} = {\rm max}\{0,\Lambda\},
\label{concurrence}
\end{equation}
where:
\begin{equation}
\Lambda=\sqrt{\lambda_{1}}-\sqrt{\lambda_{2}}-\sqrt{\lambda_{3}}-\sqrt{\lambda_{4}}\,,
\label{eq5}
\end{equation}
$\lambda_{i}$ being the eigenvalues in decreasing order of the matrix:
\begin{equation}
\rho(\sigma_{y}\otimes\sigma_{y})\rho^{*}(\sigma_{y}\otimes\sigma_{y}),
\label{eq6}
\end{equation}
where $\rho$ is the two-qubit density matrix, $\sigma_{y}$
is the second Pauli matrix and the conjugation is performed in the
computational basis. Concurrence
ranges from 0, which corresponds to a separable state, to 1 which corresponds to a maximally entangled state.

If the system is initially in the separable state $\rho_{\mathcal{S}}(0)=|eg\rangle\langle eg|$, then, by applying Eq.~(\ref{mapa2}) and tracing over the states of the environment, we end up with the following state for the system:
\begin{eqnarray}
\rho_{\mathcal{S}}(t)=&&\left(\frac{e^{-\Gamma t}+1}{2}\right)^{2}|eg\rangle\langle eg|+\left(\frac{e^{-\Gamma t}-1}{2}\right)^{2}|ge\rangle\langle ge|\nonumber\\
&&+\frac{e^{-2\Gamma t}-1}{4}(|eg\rangle\langle ge| +|ge\rangle\langle eg|)\nonumber\\
&&+\frac{1-e^{-2\Gamma t}}{2}|gg\rangle\langle gg|. \label{rhodete}
\end{eqnarray}

Simple calculation shows that the concurrence for this state is given by
\begin{eqnarray}
{\cal C}(t)=\frac{1}{2}(1-e^{-2\Gamma t})\,.\label{eq8}
\end{eqnarray}
This clearly shows that, although initially in a separable state, the system evolves to an entangled state, its concurrence reaching the maximum value of $1/2$ in the asymptotic regime $t\rightarrow\infty$. This entanglement is induced solely by an indirect interaction mediated by the common environment, since in the model considered here there is no direct interaction between the two qubits. Therefore, in order to observe the creation of entanglement it suffices to implement experimentally the second (or third) line of Eq.~(\ref{mapa2}).

The emergence of entanglement in this case can be qualitatively understood by considering that the initial separable state $|eg\rangle$  can be expressed as a superposition of the singlet $|0,0\rangle$ and triplet $|1,0\rangle$ Bell states \cite{honggangluo}:
\begin{eqnarray}
|eg\rangle=\frac{1}{\sqrt{2}}(|0,0\rangle +|1,0\rangle).
\label{eg}
\end{eqnarray} 
One can trivially see from Eq.~(\ref{lindblad}) that the singlet state does not evolve. On the other hand, the triplet state decays asymptotically to the ground state. So as time approaches to infinity we end up with the following mixed state:
\begin{equation}
\rho_{est}=
\frac{1}{2}|1,-1\rangle\langle 1,-1|+\frac{1}{2}|0,0\rangle\langle0,0|.\label{rhoest}
\end{equation}
The concurrence of the above state is easily calculated to be $1/2$, in accordance with the asymptotic limit of Eq.~(\ref{eq8}), the contribution for entanglement coming from the maximally-entangled singlet state in Eq.~(\ref{rhoest}). 

One should note that for two closely-spaced atoms,  the initial state $|eg\rangle$ is usually difficult to prepare, since it does not have the same symmetry as the ground state under exchange of particles, and the interactions with external fields are symmetrical, as the interaction with the environment. 

On the other hand, if both qubits are initially excited, i.e, $\rho_{\mathcal{S}}(0)=|ee\rangle\langle ee|$, which is in principle simpler to prepare, one should not expect to have  asymptotic entanglement, since this initial state does not have a singlet component. The question remains however whether transient entanglement might still occur.  At intermediate times, according to the first line of Eq.~(\ref{mapa1}), the state of the system becomes a mixture of the initial state, the maximally-entangled state $(|eg\rangle+|ge\rangle)/\sqrt{2}$, and the state $|gg\rangle$.  In spite of the presence of the maximally-entangled component (which eventually decays to $|gg\rangle$), the state remains separable for all times.  Indeed, the concurrence of $\rho_{\mathcal{S}}(t)$ is given by
\begin{eqnarray}
{\cal C}(t)={\rm max}\{0, {\cal C}_{1}, {\cal C}_{2}\}\label{eq8b}\,,
\end{eqnarray}
with
\begin{eqnarray}
{\cal C}_{1}(t)&=&-2e^{-2\Gamma t} \Gamma t,\nonumber\\
{\cal C}_{2}(t)&=&2e^{-2\Gamma t} (\Gamma t-\sqrt{e^{2\Gamma t}-2\Gamma t-1})\,.
\label{eq8c}
\end{eqnarray}
As ${\cal C}_{1}(t), {\cal C}_{2}(t) \leq0$, no entanglement is generated in this case. The same conclusion is obtained from Ref.~\cite{ficekspontaneous2} when the distance between the atoms goes to zero -- as shown in that reference, the dipole-dipole interaction does not play any role for this kind of initial state. 

We will thus concentrate on the state described by  Eq.~(\ref{rhodete}). Even though this state is difficult to prepare in most systems, as commented above, this is not the case in the experiment proposed here, as will be shown in  the next Section.  Using this state has an important and useful consequence. Instead of doing tomography \cite{kwiat-tomo} for full reconstruction of the state, in order to determine the entanglement through  Eq.~(\ref{concurrence}), which would require measuring sixteen observables, we can define a time-independent \textit{entanglement witness} - a unique observable that could be used to detect entanglement provided that we have some a priori knowledge about the states involved.

By definition, an entaglement witness $W$ is an observable that satisfies the following relations: ${\rm Tr}(W\rho_{sep})\geq0$ for any separable state and		${\rm Tr}(W\rho_{ent}) < 0$ for at least one entangled state, which is then said to be detected by the witness. 

For an arbitrary bipartite density matrix $\rho$, an entanglement witness is given by \cite{wu}:
\begin{eqnarray}
W=\mathbf{1}-(UV^{\dag})^{T_{A}}, \label{witness}
\end{eqnarray}
where U and V  are unitary matrices obtained from the singular value decomposition of $\rho^{T_{A}}$, $T_{A}$ being the partial transposition with respect to any one of the two qubits. The singular decomposition is given by $\rho^{T_{A}}=U\Sigma V^{\dag}$,  where $\Sigma$ is a diagonal matrix whose elements are the nonnegative square roots of the eigenvalues of $\rho^{T_{A}}{\rho^{T_{A}}}^{\dag}$ \cite{livroSDV}.

For the family of states given by Eq. (\ref{rhodete}), evaluation of Eq.~(\ref{witness}) leads in fact to a time-independent entanglement witness:
\begin{eqnarray}
W=&&\left(1+\frac{1}{\sqrt{2}}\right)|ee\rangle\langle ee|+\frac{1}{\sqrt{2}}|eg\rangle\langle ge|+\nonumber\\
&&\frac{1}{\sqrt{2}}|ge\rangle\langle eg|+\left(1-\frac{1}{\sqrt{2}}\right)|gg\rangle\langle gg|.
\label{ourwitness}
\end{eqnarray}

Furthermore, this witness is optimal for the familiy of states here considered since it leads directly to the concurrence, which can be calculated straightforwardly in terms of the mean value of $W$ as:

\begin{equation}
{\cal C}=\frac{{\rm Tr}(W\rho_{\mathcal{S}}(t))}{(1-\sqrt{2})}\,. \label{ourconcurrence}
\end{equation}
Therefore, by measuring just a single observable we are able in this case not only to detect but also to quantify entanglement, throughout the evolution of the system.

\section{EXPERIMENTAL SETUP}
\label{sec:experiment}
We discuss now the experimental setup, involving linear optics, which leads to the demonstration of environment-induced creation of entanglement, and also the setup for measuring the entanglement witness in Eq. (\ref{ourwitness}). It is possible to design a  circuit that implements the whole set of equations in Eq.~(\ref{mapa2}), but for our purposes it is enough to consider just the second line of Eq.~(\ref{mapa2}).

Our strategy for studying the dynamics of entanglement is similar to the one adopted in   Refs.~\cite{almeida07,alejo}. There, the unitary quantum map corresponding to the evolution of a single decaying atom was described in terms of the decay probability $p$, which is an exponential function of time. This allowed us to implement the decay dynamics through a static linear optics setup with a changeable parameter, namely the orientation $\theta$ of a half-wave plate, with $p=\sin^2(2\theta)$. It also allows, from the theoretical point-of-view, to encompass different kinds of dynamics in the same map, since $p$ could also be considered to be an oscillating function of time, and in this case the quantum map would correspond to Rabi oscillations of a two-level system interacting with a single mode. 

Here we generalize this approach by parametrizing the evolution in terms of two fundamental probabilities: the probability that the first qubit decays emitting one excitation into the environment; and the probability that the excitation goes from the environment to the second qubit. Different values of the coupling constant in Eq.~(\ref{eq1}) result, for the same instant of time, in different values of these probabilities, which can be tuned by changing angles $\theta_1$ of a half-wave plate and $\theta_2$ of a Dove prism in the proposed experiment. Alternatively, one may say that different values of $\Gamma t$ correspond to different values of $\theta_1$ and $\theta_2$, since the coupling constant always shows up in this combination. Of course, these two angles are not independent, since they are related through their respective dependences on $\Gamma t$. 

\subsection{SYSTEM-ENVIRONMENT EVOLUTION CIRCUIT}
\label{subsec:map circuit}

\begin{figure}[h]
\begin{center}
\includegraphics[scale=0.6]{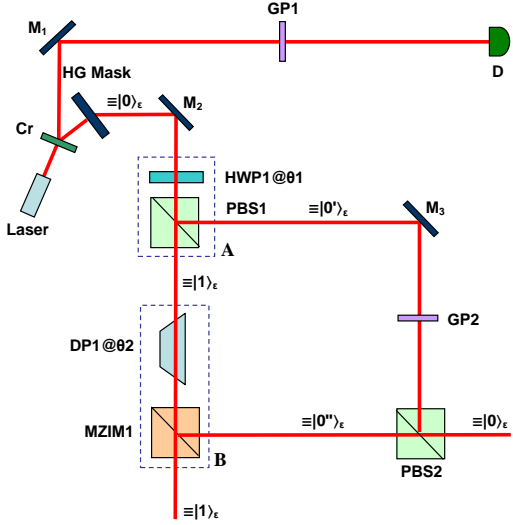}
\end{center}
\caption{\footnotesize (Color online) Optical setup to investigate
environment-induced entanglement creation. HG Mask stands for holographic mask , HWP for half-wave plate, MZIM for Mach-Zender interferometer with an additional mirror, DP for Dove prism, PBS for polarizing beam splitter, GP for glass plate, M for mirror, and D for photon detector. } \label{globalmap}
\end{figure}

Both the system and the environment are represented by different
internal degrees of freedom of a single photon. We associate the
horizontal ($H$) and vertical ($V$) polarization of the photon to the ground and
excited state of the first qubit respectively and the first-order Hermite-Gaussian
transverse modes \cite{photonics} to  the
states of the second qubit -- mode $HG_{01}$ is associated to the ground
state and $HG_{10}$ to the excited state. For simplicity, we
define the modes as: $HG_{01}\equiv |h\rangle$ and $HG_{10}\equiv |v\rangle$.
The environment states are represented by different paths of the photon. The encoding for the system is summarized in Table \ref{encoding}.
\begin{table}[t]
 \caption{\label{encoding} System encoding using polarization and Hermite-Gaussian modes.}
\begin{tabular*}{3cm}{ccc}
\cline{1-3} \vspace{-0.3cm} \\
\cline{1-3} \vspace{-0.1cm}\\
$|ee\rangle \equiv$ & $|V, HG_{10}\rangle \equiv$ & $|Vv\rangle$ \\
$|eg\rangle \equiv$ & $|V, HG_{01}\rangle \equiv$ & $|Vh\rangle$ \\
$|ge\rangle \equiv$ & $|H, HG_{10}\rangle \equiv$ & $|Hv\rangle$ \\
$|gg\rangle \equiv$ & $|H, HG_{01}\rangle \equiv$ & $|Hh\rangle$  \vspace{0.13cm}\\
\cline{1-3} \vspace{-0.3cm} \\
\cline{1-3}
\end{tabular*}
 \end{table}
Figure \ref{globalmap} shows the proposed experimental setup. It employs optical components that allow for
independent manipulation of polarization, transverse and
longitudinal (path) spatial degrees of freedom. Since in this case one has only one environment state with one excitation, $|1_{eg}\rangle_{\mathcal{E}}$ is redefined as $|1\rangle_{\mathcal{E}}$.
The circuit is composed of a spontaneous parametric down-conversion source, a holographic mask (HG Mask), a half-wave plate (HWP1), a Mach-Zender interferometer with an additional mirror (MZIM1) \cite{MZIM}, a Dove prism (DP1), two polarizing beam splitters (PBS1 and PBS2) and two glass plates (GP1 and GP2). 

\begin{figure}[h]
\begin{centering}
\includegraphics[scale=0.52]{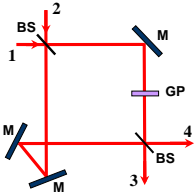}
\end{centering}
\caption{\footnotesize (Color online) Mach-Zehnder interferometer with an additional mirror. BS stands for beam splitter, GP for glass plate, and M for mirror.} \label{mzim}
\end{figure}

The MZIM is shown in Fig. \ref{mzim}. For an optical phase difference between the arms properly adjusted by tilting a glass plate (GP), $|Vh\rangle$ and $|Hv\rangle$ ($|Vv\rangle$ and $|Hh\rangle$) states entering the MZIM
through port 1 leave it from port 4(3) and the $|Vh\rangle$ and $|Hv\rangle$ states ($|Vv\rangle$ and $|Hh\rangle$)
entering through port 2 exit from port 4(3). 

The parametric down-conversion source, composed by a non-linear type-I cristal (Cr) pumped by a horizontally-polarized laser beam, produces two vertically-polarized twin photons. The two-qubit system plus environment is encoded in the lower photon. The upper photon will solely be used to validate photon counts in the lower  path in a coincidence detection regime. The glass plate GP1 is used to adjust the upper and lower optical paths.
To generate the initial state $|eg\rangle\equiv|Vh\rangle$ we place the
holographic mask, specially designed to produce the $HG_{01}$ mode,
in the path of the lower photon. The propagation direction of the
photon after the mask is associated to the vacuum state of the environment
$|0\rangle_{\mathcal{E}}$. After initial preparation in state
$|Vh\rangle_{\mathcal{S}}|0\rangle_{\mathcal{E}}$ the photon
propagates through the half-wave plate HWP1 aligned at an angle
$\theta_{1}$ with respect to the vertical polarization. The transformation
perfomed by HWP1 is:
\begin{eqnarray}
|Vh\rangle_{\mathcal{S}}|0\rangle_{\mathcal{E}}\rightarrow
\left[\cos(2\theta_{1})|Vh\rangle+\sin(2\theta_{1})|Hh\rangle\right]_{\mathcal{S}}|0\rangle_{\mathcal{E}}.
\label{phi}
\end{eqnarray}
After HWP1, the polarizing beam splitter PBS1 
reflects (transmits)  photons in the $|Vh\rangle$ ($|Hh\rangle$) state. We
associate the horizontal path after PBS1 to the vacuum state of the environment, which we call  $|0'\rangle_{\mathcal{E}}$, and the vertical path to the environment state with one
excitation $|1\rangle_{\mathcal{E}}$. This leads to:
\begin{eqnarray}
|Vh\rangle_{\mathcal{S}}|0\rangle_{\mathcal{E}}\rightarrow\cos(2\theta_{1})|Vh\rangle_{\mathcal{S}}|0'\rangle_{\mathcal{E}}+\sin(2\theta_{1})|Hh\rangle_{\mathcal{S}}|1
\rangle_{\mathcal{E}}.\nonumber\\
\label{phidois}
\end{eqnarray}
The physical interpretation of the transformation (\ref{phidois}), performed by the dashed circuit block A,  is that there is a probability amplitude $\cos(2\theta_{1})$ that the system and environment remain in the same state $|eg\rangle_{\mathcal S}|0\rangle_{\mathcal E}$, and a probability amplitude $\sin(2\theta_{1})$ that the first qubit decays emitting one excitation into the environment.   

In the path associated to the state of the environment with one excitation $|1
\rangle_{\mathcal{E}}$, the photon
propagates through a Dove prism rotated at an
angle $\theta_{2}$, that accomplishes the following transformation: 
\begin{equation}
|Hh\rangle_{\mathcal{S}}|1
\rangle_{\mathcal{E}}\rightarrow\left[\cos(2\theta_{2})|Hh\rangle+
\sin(2\theta_{2})|Hv\rangle\right]_{\mathcal{S}}|1\rangle_{\mathcal{E}}.
\label{phitres}
\end{equation}
After the Dove prism DP1, the Mach-Zender interferometer MZIM1
conserves the path of the $|Hh\rangle$ state and changes by $90^{\circ}$ the path of the $|Hv\rangle$ state. We
associate the horizontal path to the vacuum state of the environment which we call  $|0''\rangle_{\mathcal{E}}$ and the vertical path after the MZIM1 to the environment state with one
excitation $|1\rangle_{\mathcal{E}}$. The resulting transformation after the MZIM1 is the following:
\begin{eqnarray}
|Hh\rangle_{\mathcal{S}}|1
\rangle_{\mathcal{E}}\rightarrow\cos(2\theta_{2})|Hh\rangle_{\mathcal{S}}|1\rangle_{\mathcal{E}}+
\sin(2\theta_{2})|Hv\rangle_{\mathcal{S}}|0''\rangle_{\mathcal{E}}.\nonumber\\
\label{phiquatro}
\end{eqnarray}
The physical interpretation of the transformation (\ref{phiquatro}), performed by  the dashed circuit block B,  is that there is a probability amplitude $\cos(2\theta_{2})$ that the system and environment remain in the same state, corresponding to the two qubits in the state $|g\rangle$ and one excitation in the environment, and a probability amplitude $\sin(2\theta_{2})$ that the excitation goes from the environment to the second qubit.

Finally, the polarizing beam splitter PBS2 together with the GP2 allows for the in-phase coherent combination of the two paths associated with the vacuum state of the environment $|0'\rangle_{\mathcal{E}}$ and $|0''\rangle_{\mathcal{E}}$ into a single horizontal path $|0\rangle_{\mathcal{E}}$ as shown in Figure \ref{globalmap}. Therefore the overall transformation of the whole circuit is:
\begin{eqnarray}
|Vh\rangle_{\mathcal{S}}|0\rangle_{\mathcal{E}}\rightarrow&&\big[\cos(2\theta_{1})|Vh\rangle\nonumber\\
&&+\sin(2\theta_{1})\sin(2\theta_{2})|Hv\rangle\big]_{\mathcal{S}}|0\rangle_{\mathcal{E}}\nonumber\\
&&+\sin(2\theta_{1})\cos(2\theta_{2})|Hh\rangle_{\mathcal{S}}|1\rangle_{\mathcal{E}}\,.
\label{overall}
\end{eqnarray}

This equation coincides precisely with the second line of Eq.~(\ref{mapa2}), upon associating
$\cos(2\theta_{1})$, $\sin(2\theta_{1})\sin(2\theta_{2})$
and $\sin(2\theta_{1})\cos(2\theta_{2})$ with the coefficients $Q$, $R$, and $S$ in that equation, respectively. 

Note that the coefficient $Q$ is the probability amplitude that the first qubit remains in the excited state, $R$ is the probability amplitude that the first qubit decays, emitting one excitation into the environment, multiplied by the probability amplitude that the second qubit absorbs the excitation present in the environment, and $S$ is the probability amplitude  that the first qubit decays multiplied by the probability amplitude that the second qubit remains unexcited, so that the excitation emitted by the first qubit remains in the environment.

The interaction of the system with the environment is thus achieved in a entirely controllable manner: each value of $\Gamma t$ corresponds to a specific set of angles $\theta_1$ and $\theta_2$.  Furthermore, this setup makes it clear that the interaction between the two qubits is  mediated by the environment. One should also note that the present proposal allows for the easy preparation of the initial state $|eg\rangle$, which, as mentioned before, may be a demanding task in other systems. 

\subsection{MEASUREMENT CIRCUIT}
\label{subsec:measurement circuit}

In order to measure the mean value of the entanglement witness (\ref{ourwitness}) we decompose it in its diagonal basis, which happens to be the collective basis:

\begin{eqnarray}
W&=&\left(1+\frac{1}{\sqrt{2}}\right)|1,1\rangle\langle 1,1|+\frac{1}{\sqrt{2}}|1,0\rangle\langle1,0|-\nonumber\\
&&\frac{1}{\sqrt{2}}|0,0\rangle\langle0,0|+\left(1-\frac{1}{\sqrt{2}}\right)|1,-1\rangle\langle 1,-1|.\nonumber\\
\label{diagonalwitness}
\end{eqnarray}

\begin{figure}[h]
\begin{center}
\includegraphics[scale=0.65]{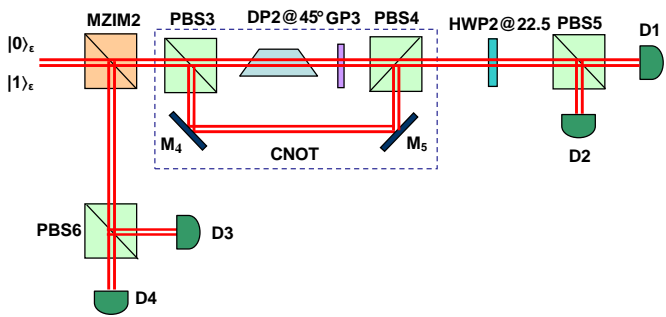}
\end{center}
\caption{\footnotesize  (Color online) Optical setup to measure the entanglement witness. MZIM stands for Mach-Zender interferometer with an additional mirror, PBS for polarizing beam splitter, DP for Dove prism, HWP for half-wave plate, GP for glass plate, M for mirror, and D for photon detector.} 
\label{globalmeasure}
\end{figure}

In Fig. \ref{globalmeasure} we show the circuit that performs projective measurements of $\rho_{\mathcal{S}}(t)$ in the collective basis. Photons in modes $|0\rangle_{\mathcal{E}}$ and $|1\rangle_{\mathcal{E}}$ outgoing the system-environment evolution circuit are incoherently driven to the measurement circuit, which corresponds to a trace over the states of the environment.

The circuit is composed of a Mach-Zender interferometer with an additional mirror (MZIM2), two polarizing beam splitters (PBS5 and PBS6), a CNOT gate~\cite{oliveira05}, a half-wave plate (HWP2), and four photon detectors (D1, D2, D3 and D4).

The CNOT gate is an interferometer composed by two polarizing beam
splitters (PBS3 and PBS4), a Dove
Prism (DP2) and a glass plate (GP3). PBS3 splits horizontal (upper path) and vertical
polarized photons (lower path). Rotated at $45^{\circ}$ the Dove prism
in the upper path flips between Hermite-Gaussian modes:
$HG_{01}\leftrightarrow HG_{10}\equiv h\leftrightarrow v$. The optical lengths of the upper and lower paths are adjusted by the glass plate GP3 enabling coherent  in-phase combination in PBS4. This
arrangement corresponds to the usual CNOT gate where the polarization degree
of freedom acts as the control bit and the transverse mode acts as
the target bit, according to Table \ref{CNOT}.
\begin{table}[h]
 \caption{\label{CNOT} CNOT gate. C stands for control and T for target.}
 \begin{tabular*}{3cm}{cccc}
\cline{1-4} \vspace{-0.3cm} \\
\cline{1-4} \vspace{-0.1cm}\\
Input & & & Output \\
C  T & & & C  T \\
$|V  v\rangle$ &   &  &$|V  v\rangle$ \\
$|V  h\rangle$ &   &  &$|V  h\rangle$ \\
$|H  v\rangle$ &   &  &$|H  h\rangle$ \\
$|H  h\rangle$ &   &  &$|H  v\rangle$ \vspace{0.13cm}\\
\cline{1-4} \vspace{-0.3cm} \\
\cline{1-4}
\end{tabular*}
 \end{table}

The MZIM2 conserves the path of the $|Vh\rangle$ and $|Hv\rangle$ states and changes by $90^{\circ}$ the path of the $|Vv\rangle$ and $|Hh\rangle$  states.
This implies that photons in states $|1,0\rangle$ and $|0,0\rangle$ will be transmitted by the MZIM2 while photons in states $|1,1\rangle$ and $|1,-1\rangle$ will be reflected by it.

In the horizontal path after the MZIM2, the CNOT gate circumscribed in the dashed block and described by Table \ref{CNOT}, plus HWP2 aligned at $22.5^{\circ}$ with respect to the horizontal polarization, which plays the role of a Hadamard gate in the polarization degree of freedom, i.e.,
$|H\rangle\rightarrow\frac{1}{\sqrt{2}}(|H\rangle+|V\rangle)$ and $|V\rangle\rightarrow\frac{1}{\sqrt{2}}(|H\rangle-|V\rangle$, performs the following transformations: $|1,0\rangle\rightarrow|Hh\rangle$ and
$|0,0\rangle\rightarrow-|Vh\rangle$. Then, by virtue of PBS5, which transmits the horizontal polarization and reflects the vertical one, D1 and D2 detect the populations of $|1,0\rangle$ and
$|0,0\rangle$, respectively. In the vertical path after the MZIM2, due to PBS6, populations of $|1,1\rangle$ and $|1,-1\rangle$ are detected in D3 and D4, respectively.

Let $C_{i}$ be the photon counts in detector $i$ and $N=\sum_{i=1}^{4} C_{i}$ the total counts during the measurement interval. Then, according to Eq. (\ref{diagonalwitness}),
\begin{eqnarray}
Tr(W\rho_{\mathcal{S}}(t))=&\frac{1}{N} \Big[ C_{1}\frac{1}{\sqrt{2}}+C_{2}\left(-\frac{1}{\sqrt{2}}\right)+C_{3}
\left(1+\frac{1}{\sqrt{2}} \right) \nonumber\\
&+C_{4} \left( 1-\frac{1}{\sqrt{2}} \right)\Big].
\label{wmeanvalue}
\end{eqnarray}

The concurrence follows straightforwardly through Eq. (\ref{ourconcurrence}).

\section{SINGLE-MODE ENVIRONMENT AND INITIAL ENTANGLED STATES}
\label{sec:generalization of the model}

Our proposal can be easily generalized to describe the evolution of other initial states as well as the oscillatory exchange of energy between the two-qubit system and a single mode of the environment. As mentioned in Section \ref{sec:experiment}, the approach of unitary quantum maps  is quite general and accommodates other  types of dynamics, depending on the way we parametrize the coefficients with time. Indeed, Eq.~(\ref{mapa2}) can match the non-dissipative dynamics of two non-interacting qubits in a resonant cavity, described by the following Hamiltonian:
\begin{equation}
H= \sum_{i=1}^2 \hbar \omega S_{i}^{z}  +
\hbar \omega(a^{\dag}a+1/2)
 - i\hbar g
\sum_{i=1}^2\left((S_{i}^{+}a-S_{i}^{-}a^\dagger\right).
\label{hcavity}
\end{equation}
where ${g}$ is the coupling constant, while $a$ and $a^{\dag}$ are the annihilation and creation operators of the cavity field mode. In this situation $|1_{ee}\rangle_{\mathcal{E}}$ and $|1_{eg}\rangle_{\mathcal{E}}$ are identified, since there is just a single mode in the cavity. No major changes, except for the time dependence of the coefficients, are needed to represent this rather different dynamics. As a matter of fact, the circuit describing the dissipative dynamics can be applied to the resonant case as long as the coefficients of the map in Eq. (\ref{mapa2})  are given as functions of time by:
$M= \frac{1}{2}(1+\cos{(2gt)})$, $P=-\frac{i}{2}\sin{(2gt)}$, $N=\frac{1}{2}(-1+\cos{(2gt)})$, $Q= \frac{1}{2}(1+\cos{(\sqrt{2}gt)})$, $R=\frac{1}{2}(-1+\cos{(\sqrt{2}gt)})$, $S= \frac{-i}{\sqrt{2}}{\sin{(\sqrt{2}gt)}}$. The extra phase $i$ can be easily compensated by a phase plate.

The angles $\theta_1$ and $\theta_2$ in the proposed experimental setup are now functions of $gt$. Tuning the coupling constant, or rather the product $gt$, corresponds to tuning those two angles, according to the above equations. 

Our setup encompasses therefore both the decay and the oscillatory case, as in the experiment reported in Ref.~\cite{almeida07}. This is a great advantage of our parametrization of the evolution of the system.

Similar circuits can be applied to analyze the dynamics of initially entangled states, which are easily prepared within the present framework. For instance, the state $\alpha|eg\rangle+\beta|ge\rangle$ is obtained from the initial state  $|eg\rangle$ by applying a half-wave plate to the polarization mode and the CNOT circuit in Fig.~\ref{globalmeasure} in order to entangle the two degrees of freedom. 

The dynamics of the family of states $\alpha|gg\rangle+\beta|ee\rangle$ under the action of a global environment is particularly rich, giving rise not only to the disappearance of entanglememt at finite times but also to entanglement revival. A detailed analysis was presented in Ref.~\cite{ficekdark}, where it was shown that in this case the evolution of the density matrix elements is independent of the dipole-dipole interaction between the atoms. Therefore, the results obtained from our model precisely coincide with those in Ref.~\cite{ficekdark}, in the limit when the distance between the two atoms goes to zero.

\section{CONCLUSIONS}
\label{sec:conclusion}

In this work we propose a simple experiment, which uses only elementary linear optical devices, to observe creation of entanglement between two non-interacting qubits, induced by a common environment. 

We encode the two qubits and the environment into different degrees of freedom of a single photon: the polarization, the transverse modes, and the different paths stemming from different directions of the photon momentum.  Our experimental setup  directly implements the unitary map that describes the time evolution of the system and the environment. We recover the dynamics of the two-qubit system by tracing over the path of the photon, which represents the environment. 

The evolution map is parametrized in terms of time-dependent transition probabilities, rather than time itself. This allows the investigation of the dynamics of entanglement through a static setup. It also follows from this parametrization that the same setup describes not only the collective interaction of two qubits with a zero-temperature bath, but also the reversible exchange of excitation between the two qubits and a single cavity mode.

Entanglement is quantified by a time-independent entanglement witness -- a single observable -- shown to be proportional to the concurrence for the  states here considered. Therefore, our proposed scheme to quantify entanglement does not require time-consuming tomographic measurements to determine the state of the system.  

\section{Appendix}
\label{sec:apendice}

\subsection{Kraus Operators}
\label{Kraus}

The first step to obtain the Kraus operators is to determine $\rho_{\mathcal{S}}(t)$ solving the Lindblad equation (\ref{lindblad}). In the collective basis $\{|0,0\rangle\equiv\frac{1}{\sqrt{2}}(|eg\rangle-|ge\rangle), |1,1\rangle\equiv|ee\rangle, |1,0\rangle\equiv \frac{1}{\sqrt{2}}(|eg\rangle+|ge\rangle), |1,-1\rangle\equiv|gg\rangle\}$, which we denote by the indices $1,2,3,4$, respectively, $\rho_{\mathcal{S}}(t)$ has the following elements:
\begin{eqnarray}
\rho_{11}(t)&=&\rho_{11}(0),\nonumber\\
\rho_{12}(t)&=&\rho_{12}(0)e^{-\Gamma t},\nonumber\\ 
\rho_{13}(t)&=& \rho_{13}(0)e^{-\Gamma t},\nonumber\\ 
\rho_{14}(t)&=&\rho_{14}(0),\nonumber\\ 
\rho_{22}(t)&=&\rho_{22}(0)e^{-2\Gamma t},\nonumber\\
\rho_{23}(t)&=&\rho_{23}(0)e^{-2\Gamma t},\nonumber\\ 
\rho_{24}(t)&=&\rho_{24}(0)e^{-\Gamma t},\nonumber\\ 
\rho_{33}(t)&=&\rho_{33}(0)e^{-2\Gamma t}+2\rho_{22}(0)\Gamma t e^{-2\Gamma t},\nonumber\\ 
\rho_{34}(t)&=& \rho_{34}(0)e^{-\Gamma t}+2\rho_{23}(0)e^{-\Gamma t}(1-e^{-\Gamma t}),\nonumber\\
\rho_{44}(t)&=&\rho_{44}(0)+\rho_{22}(0)(1-e^{-2\Gamma t}-2\Gamma t e^{-2\Gamma t})\nonumber\\
&&+\rho_{33}(0)(1-e^{-2\Gamma t}).
\label{collectiveevolution}
\end{eqnarray}

We define now the matrix $E^{ij}(t)$ as $\rho_{\mathcal{S}}(t)$ with the initial conditions $\rho_{mn}(0)=\delta_{mi}\delta_{nj}$.  For example, $E^{22}(t)$ is given by $\rho_{\mathcal{S}}(t)$ considering solely $\rho_{22}(0)=1$ and all other coefficients equal to zero. Using (\ref{collectiveevolution}) we get:
\begin{eqnarray}
E^{22}(t)=\left(\begin{array}{cccc}
           0&0&0&0\\
           0&e^{-2\Gamma t}&0&0\\
            0&0&2\Gamma t e^{-2\Gamma t}&0\\
            0&0&0&0
           \end{array}\right)\,.
           \label{example}
\end{eqnarray}

 The next step is to build the $16\times16$ positive semidefinite Choi matrix \cite{havel} composed of matrices $E^{ij}(t)$ as follows: 
\begin{eqnarray}
C=\left(\begin{array}{cccc}
           E^{11}(t)&E^{12}(t)&\cdots\\
           E^{21}(t)&E^{22}(t)&\cdots\\
           \vdots&\vdots&\ddots
          \end{array}\right).
           \label{choimatrix}
\end{eqnarray}

Matrix C can be represented by:

\begin{eqnarray}
C=\sum_{\mu}a_{\mu}|a_{\mu}\rangle\langle a_{\mu}|\equiv\sum_{\mu}|v_{\mu}\rangle \langle v_{\mu}|,
 \label{choidecomposition}
\end{eqnarray}
where $|a_{\mu}\rangle$ are the normalized eigenvectors of C with eigenvalues $a_{\mu}$, and $|v_\mu\rangle$ are non-normalized vectors. In the present case we have just four non-null vectors $|v_{\mu}\rangle$, each one corresponding to a Kraus operator. For convenience we designate $\mu=0,1_{A},1_{B},2$.

The final step is to divide each column matrix corresponding to $|v_{\mu}\rangle$ into four segments of equal length. Then the matrix representing Kraus operator $M_{\mu}$ has the $i$th segment of $|v_{\mu}\rangle$ as its $i$th  column. 

Following this procedure, we end up with the following Kraus operators:
\begin{eqnarray}
M_{0}=\left(\begin{array}{cccc}
           1&0&0&0\\
           0&A&0&0\\
            0&0&A&0\\
            0&0&0&1
           \end{array}\right),
           \label{kraus0}
\end{eqnarray}

\begin{eqnarray}
M_{1A}=\left(\begin{array}{cccc}
           0&0&0&0\\
           0&0&0&0\\
            0&B&0&0\\
            0&0&C&0
           \end{array}\right),
           \label{kraus1A}
\end{eqnarray}

\begin{eqnarray}
M_{1B}=\left(\begin{array}{cccc}
           0&0&0&0\\
           0&0&0&0\\
            0&D&0&0\\
            0&0&E&0
           \end{array}\right),
           \label{kraus1B}
\end{eqnarray}
\begin{eqnarray}
M_{2}=\left(\begin{array}{cccc}
           0&0&0&0\\
           0&0&0&0\\
            0&0&0&0\\
            0&F&0&0
           \end{array}\right),
           \label{kraus2}
\end{eqnarray}
where: \vspace{0.2cm} \\
$A=e^{-\Gamma t}$,
$B=\frac{\alpha_{1}\alpha_{2}}{\sqrt{2}\gamma_{1}}$,
$C=\frac{\alpha_{2}}{\sqrt{2}\gamma_{1}}$,
$D=\frac{\beta_{1}\beta_{2}}{\sqrt{2}\gamma_{2}}$,
$E=\frac{\beta_{2}}{\sqrt{2}\gamma_{2}}$, \vspace{0.1cm}\\
$F=\sqrt{1-e^{-2\Gamma t}-2e^{-2\Gamma t}-\Gamma t}$,\vspace{0.1cm}\\
$\Omega=\sqrt{17-32e^{\Gamma t}+e^{4\Gamma t}+e^{2\Gamma t}(14-4\Gamma t)+4\Gamma t(1+\Gamma t)}$, \vspace{0.1cm}\\
$\alpha_{1}=\frac{1-e^{2\Gamma t}+2\Gamma t-\Omega}{4(-1+e^{\Gamma t})}$,\vspace{0.1cm}\\ 
$\beta_{1}=\frac{1-e^{2\Gamma t}+2\Gamma t+\Omega}{4(-1+e^{\Gamma t})}$,\vspace{0.1cm}\\
$\alpha_{2}=\sqrt{e^{-2\Gamma t}(-1+e^{2\Gamma t}+2\Gamma t-\Omega)}$, \vspace{0.1cm}\\
$\beta_{2}=\sqrt{e^{-2\Gamma t}(-1+e^{2\Gamma t}+2\Gamma t+\Omega)}$,\vspace{0.1cm}\\
$\gamma_{1}=\sqrt{1+\frac{(-1+e^{2\Gamma t}-2\Gamma t+\Omega)^{2}}{16(-1+e^{\Gamma t})^{2}}}$,\vspace{0.1cm}\\
$\gamma_{2}=\sqrt{1+\frac{(1-e^{2\Gamma t}+2\Gamma t+\Omega)^{2}}{16(-1+e^{\Gamma t})^{2}}}$.\vspace{0.1cm}\\

\subsection{Numeric Coefficients}
\label{coef}

Here we show the time-dependent coefficients appearing in  Eq. (\ref{eq16}):\\

\noindent
$X=\frac{\delta_{1}(1-e^{2\Gamma t}+2\Gamma t-\Omega)}{\epsilon_{1}8(-1+e^{\Gamma t})}$,
$Y=\frac{(1-e^{2\Gamma t}+2\Gamma t+\Omega)\delta_{2}}{8(-1+e^{\Gamma t})\epsilon_{2}}$,
$Z=\frac{\delta_{1}}{2\epsilon_{1}}$,
$W=\frac{\delta_{2}}{2\epsilon_{2}}$,\\
$\delta_{1}=\sqrt{e^{-2\Gamma t}(-1+e^{2\Gamma t}+2\Gamma t-\Omega)}$,\\ $\epsilon_{1}=\sqrt{1+\frac{(-1+e^{2\Gamma t}-2 \Gamma t+\Omega)^2}{16(-1+e^{\Gamma t})^2}}$,\\
$\delta_{2}=\sqrt{e^{-2\Gamma t}(-1+e^{2\Gamma t}+2\Gamma t+\Omega)}$, \\
$\epsilon_{2}=\sqrt{1+\frac{(1-e^{2\Gamma t}+2 \Gamma t+\Omega)^2}{16(-1+e^{\Gamma t})^2}}$.\\

\begin{acknowledgments}
We would like to thank  Fernando de Melo, Leandro Aolita, Marcelo Fran\c ca Santos, Nicim Zagury, Ruynet Lima de Matos Filho, and Stephen P. Walborn for discussions during the development of this work. The authors acknowledge financial support from the Brazilian funding agencies CNPq, CAPES and FAPERJ.  This work was performed as part of the Brazilian Millennium Institute for Quantum Information.
\end{acknowledgments}

\end{document}